\documentclass[a4paper,12pt]{article}
\pdfoutput=1
\usepackage[utf8]{inputenc}
\usepackage{graphicx}
\usepackage{caption}
\usepackage{subcaption}
\usepackage{amsmath,amssymb}
\usepackage[margin=0.8in]{geometry}
\usepackage{color}
\usepackage[colorlinks=true,linkcolor=blue,citecolor=blue]{hyperref}
\usepackage{float}
\usepackage{comment}
\usepackage[usenames,dvipsnames]{xcolor}
\usepackage[ruled,vlined]{algorithm2e}
\usepackage{authblk}

\title{A threshold-force model for adhesion and mode I fracture}
\author[1]{Srivatsan Hulikal\thanks{srivatsan\_hulikal@brown.edu}}
\author[2]{Kaushik Bhattacharya\thanks{bhatta@caltech.edu}}
\author[2]{Nadia Lapusta\thanks{lapusta@caltech.edu}}
\affil[1]{School of Engineering, Brown University, Providence, RI 02912}
\affil[2]{Department of Mechanical and Civil Engineering, California Institute of Technology, Pasadena, CA 91125}

\date{}

\begin{document}

\maketitle

\begin{abstract} 

We study the relation between a threshold-force based model at the microscopic
scale and mode I fracture at the macroscopic scale in a system of discrete
interacting springs. Specifically, we idealize the contact between two surfaces
as that between a rigid surface and a collection of springs with long-range
interaction and a constant tensile threshold force.  We show that a particular
scaling similar to that of crack-tip stress in Linear Elastic Fracture
Mechanics leads to a macroscopic limit behavior. The model also reproduces the
scaling behaviors of the JKR model of adhesive contact. We determine how the
threshold force depends on the fracture energy and elastic properties of the
material. The model can be used to study rough-surface adhesion.

\end{abstract}

\section{Introduction}

Adhesion plays an important role in many phenomena, especially at small
length-scales where surface area to volume ratio is large
\cite{maboudian1997critical,geim2003microfabricated}. The JKR model
\cite{johnsonKL:1} has been widely used to study adhesion
\cite{Fuller1975a,Horn1987,Chu2005}. Here, we show that, in a system of
interacting springs, a tensile threshold-force criterion for each spring at the
microscopic scale gives rise to fracture mechanics and JKR at a larger scale. 

Specifically, we represent a surface using a system of discrete interacting
springs with a tensile threshold force and use it to study the adhesive contact
of surfaces. We find that a particular scaling of the threshold force with
discretization size is necessary to make the macroscopic response independent
of discretization. Remarkably, this scaling is exactly the same as that of
crack-tip stress in Linear Elastic Fracture Mechanics ($1/\sqrt{r}$, where $r$
is the distance from the crack tip). The JKR adhesion model can also be seen
from the perspective of fracture mechanics \cite{maugis1992adhesion}. Using our
model, we study the contact of a sphere with a rigid flat surface. The model
reproduces two scalings of the JKR theory, the linear dependence of the maximum
tensile force on fracture energy and radius of the sphere. This helps determine
the threshold force in our model as a function of the fracture energy and
elastic properties of the material. 

The model formulation, algorithm for numerical implementation, and a validation
through simulation of Hertzian contact are described in Section
\ref{sec:formulation}. Sections \ref{sec:adhesiveContact} and \ref{sec:JKR}
present the connection to fracture mechanics and JKR theory. We conclude with
suggestions for applications and further development of the model in Section
\ref{sec:conclusion}.

\section{Formulation}\label{sec:formulation}

\begin{figure}[H]
\centering
\makebox[\textwidth][c]{\includegraphics[scale=0.4]{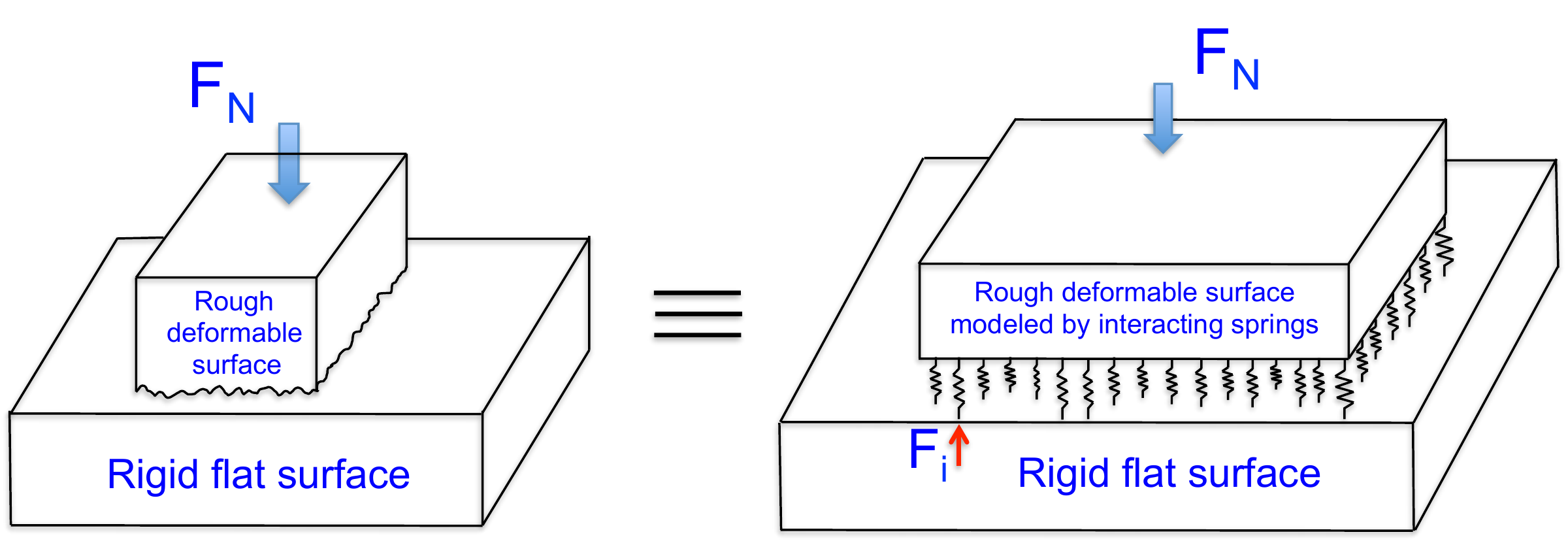}}
\caption{Contact of a rigid flat surface and a rough deformable one with the
deformable surface modeled using a set of interacting springs. The only degree
of freedom of the springs is normal to the interface.}
\label{fig:systemDiscretizationChapterCriticalTensileForceModel}
\end{figure}

Consider the contact of two surfaces, one rough and deformable, the other flat
and rigid. We represent the deformable surface with a set of springs as shown
in Figure \ref{fig:systemDiscretizationChapterCriticalTensileForceModel}. 
Constitutive equations relate the deformations and forces of these springs.
We prescribe that each spring in contact can sustain a maximum
tensile force $F_{th}$. If the force on the spring exceeds $F_{th}$, it snaps
out of contact and the force becomes zero. The governing equations are: 
\begin{equation}\label{eq:governingEquations}
u_i = C_{ij} F_j, \quad F_i \leq F_{th}, \quad x_i \leq d,
\end{equation}
where $u_i$ is the length change of the spring `i', $F_j$ is the force on spring
`j', $C_{ij}$ is the compliance that captures the effect of the force at `j' on
an element at `i', $F_{th}$ is the threshold force, $x_i = x^0_i + u_i $ is the
length of the spring `i' with undeformed length $x^0_i$, and dilatation $d$
is the distance between the two surfaces. The third expression in
(\ref{eq:governingEquations}) is a kinematic constraint corresponding to the
rigidity of the flat surface. The undeformed lengths $x^0_i$ can be used to
approximate a desired surface geometry.  The compliance $C_{ij}$ is derived
from the superposition of the Boussinesq solution for a point force in a
semi-infinite half space \cite{johnsonKL_book}, 
\begin{equation}\label{eq:complianceBoussinesq}
C_{ij} = \begin{cases}
\frac{1-\nu}{2\pi G} \frac{1}{r_{ij}} & i \neq j,\\
\frac{1-\nu}{2\pi G} \frac{3.8}{\Delta} & i = j,
\end{cases}
\end{equation}
where $\nu,G$ are the Poisson's ratio and shear modulus of the material, and
$r_{ij}$ is the distance between springs `i' and `j', and $\Delta$ is the
distance between two neighboring springs. $1/r_{ij}$ becomes singular at
$i=j$ since $r_{ij} = 0$. This singularity is regularized by assuming that
$F_i$ is uniformly distributed over a square area with side $\Delta$. The
displacements caused by such a pressure distribution was derived by Love
\cite{love_AEH:1}. This solution is very close to the Boussinesq solution even
for neighboring springs but it is not singular for $i=j$. The case $i=j$ in
Equation (\ref{eq:complianceBoussinesq}) means that a force at spring `i' deforms
it 3.8 times more than a neighboring one. Thus, in computing $u_i$ due to
$F_j$, we use the Boussinesq solution for all $i \neq j$ and the Love solution
for $i = j$. A brute-force computation of the $1/r_{ij}$ kernel becomes
expensive for large systems. We use the Fast Multipole Method
\cite{greengard1987} to do this calculation faster. Similar models have been
used to study contact, for example see \cite{bora_CK:1} and references therein.
Our contribution is the extension to adhesion.

Nondimensionalizing (\ref{eq:governingEquations}) and
(\ref{eq:complianceBoussinesq}) using $L^*$ for length and $F^*$ for force, we
get:
\begin{equation}\label{eq:governingEquationsNondim}
\bar{u}_i = \bar{C}_{ij} \bar{F}_j, \quad \bar{F}_i \leq \bar{F}_{th},
\quad \bar{x}_i \leq \bar{d}.
\end{equation}
Setting $F^* = 2\pi G{L^*}^2/(1-\nu)$, 
\begin{equation} \label{eq:complianceBoussinesqNondim}
\bar{C}_{ij} = \begin{cases}
1/\bar{r}_{ij} & i \neq j,\\
3.8/\bar{\Delta} & i = j.
\end{cases}
\end{equation}

\subsection{Algorithm}\label{sec:algorithm}

The forces and the deformations of the springs are coupled through the elastic
interactions. Thus, when a spring snaps out of contact, it changes the forces
and deformations of neighboring and distant springs (which might then exceed
their threshold force). Here, we describe the algorithm we use to simulate
displacement-controlled (prescribed dilatation) loading/unloading of the
surface.

\begin{algorithm}[h] \label{alg:contact}
Given current state $\bar{u}_i(\bar{t}), \bar{F}_i(\bar{t}), \bar{d}(\bar{t})$ and
	the set of springs in contact $\mathcal{I}_c$ \\
\Repeat ( Take a time step $\bar{t} \to \bar{t}+\Delta \bar{t},
\bar{d}^{\text{new}} \to \bar{d} + \Delta \bar{d}$ ){$\bar{t}=\bar{T}$}
{
\Repeat{All conditions in (\ref{eq:governingEquationsNondim}) are met}
{		1. \For{$i=1, \dots, N$}{
			\textbf{if} $i \in \mathcal{I}_c$, set $\bar{u}_i =
\bar{d}^{\text{new}}-\bar{x}^0_i$; 
			\textbf{else}, set $\bar{F}_i = 0$ \\
		}
			compute $\bar{F}_i$ for $i \in \mathcal{I}_c$ and $\bar{u}_i$ for $i
\notin \mathcal{I}_c$ using the equation in (\ref{eq:governingEquationsNondim})
and Equation (\ref{eq:complianceBoussinesqNondim}).\\
		2. \For{$i=1, \dots, N$}{
			\textbf{if} $\bar{F}_i > \bar{F}_{th}$, set $\bar{F}_i = 0$, remove $i$ from $\mathcal{I}_c$ \\
			\textbf{if} $\bar{x}^0_i + \bar{u}_i > \bar{d}^{\text{new}}$, add $i$ to $\mathcal{I}_c$ \\
\textbf{if} one or both of above is true, go back to Step 1.
		}
}
}
\caption{Algorithm to simulate evolution of the spring system for
displacement-controlled loading/unloading.}
\end{algorithm}

\subsection{Validation using Hertzian contact} 
\label{subsec:hertzianContactValidation}

To validate our formulation, we simulate Hertzian contact of a linear-elastic
sphere of radius $1$ against a rigid flat surface.  The two surfaces are
initially apart, and the forces and deformations of all the springs are zero.
We then decrease the dilatation and compute the evolution of the forces and
deformations of the springs using the algorithm described above. The total
macroscopic force is obtained as the sum of the spring forces. Compressive
values of the force are shown as positive.

\begin{figure}[H]
\centering
\begin{subfigure}[t]{0.49\textwidth}
  \includegraphics[width=\textwidth]{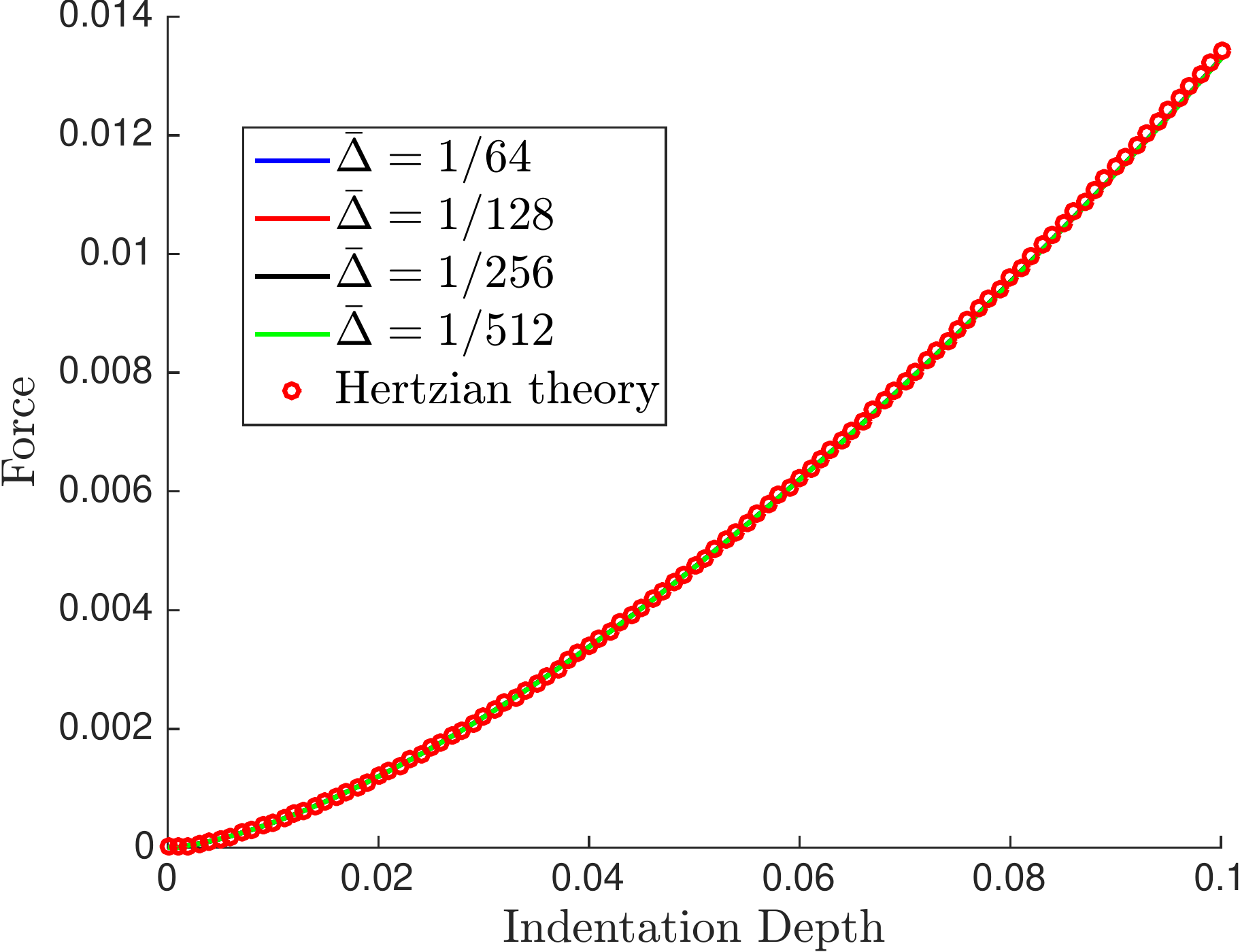}
\subcaption{}
\label{fig:dilatationForceHertzianContact}
\end{subfigure} 
\begin{subfigure}[t]{0.49\textwidth}
  \includegraphics[width=\textwidth]{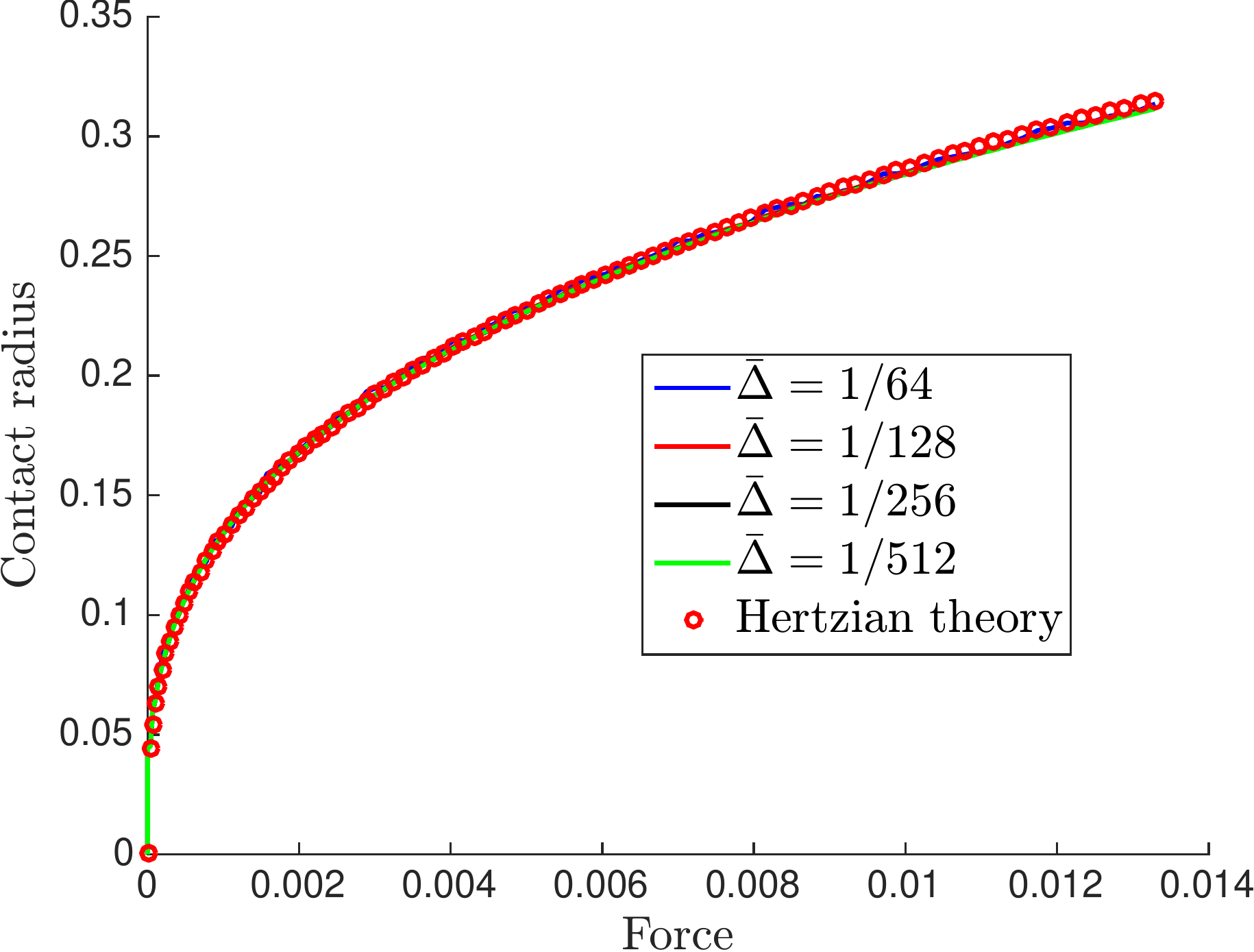}
\subcaption{}
\label{fig:forceContactRadiusHertzianContact}
\end{subfigure} 
\caption{Comparison of analytical and numerical solutions (at different
discretizations) for contact of a linear-elastic sphere against a rigid flat
surface (a) indentation depth vs force, (b) force vs contact radius. The model
shows a good match with the analytical results.}
\label{fig:hertzianContactValidation}
\end{figure}

The model shows a surprisingly good match with the analytical solution (Figure
\ref{fig:hertzianContactValidation}). The elastic constants used in the
analytical and numerical solutions are the same and no other parameters have
been used in obtaining the numerical results.

\section{Adhesive contact} \label{sec:adhesiveContact}

Let us now move to the adhesive contact of a sphere. The two surfaces are
initially apart and the forces and deformations of all the springs are zero. We
then do a displacement-controlled loading-unloading test. We first compress the
two surfaces into contact (loading) and then pull them apart (unloading).  The
parameters used in this simulation are: radius of sphere = 1, $\bar{\Delta} =
1/128, \bar{F}_{th} = 2\times 10^{-5}$.

\begin{figure}[H]
\centering
\begin{subfigure}[t]{0.49\textwidth}
\includegraphics[width=\textwidth]{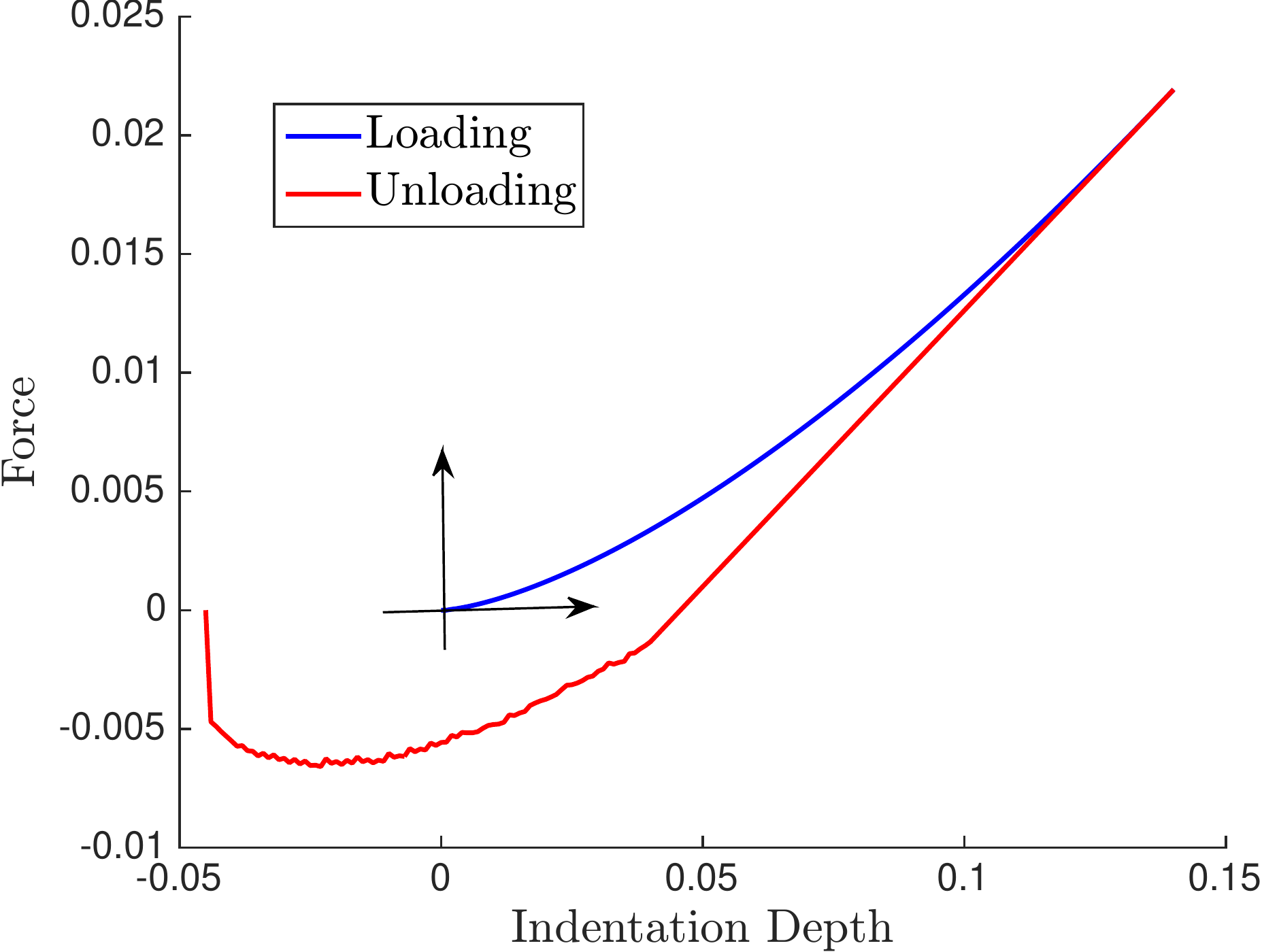}
\subcaption{}
\label{fig:dilatationForceLoadingUnloading128by128ScalingPower0}
\end{subfigure} 
\begin{subfigure}[t]{0.49\textwidth}
\includegraphics[width=\textwidth]{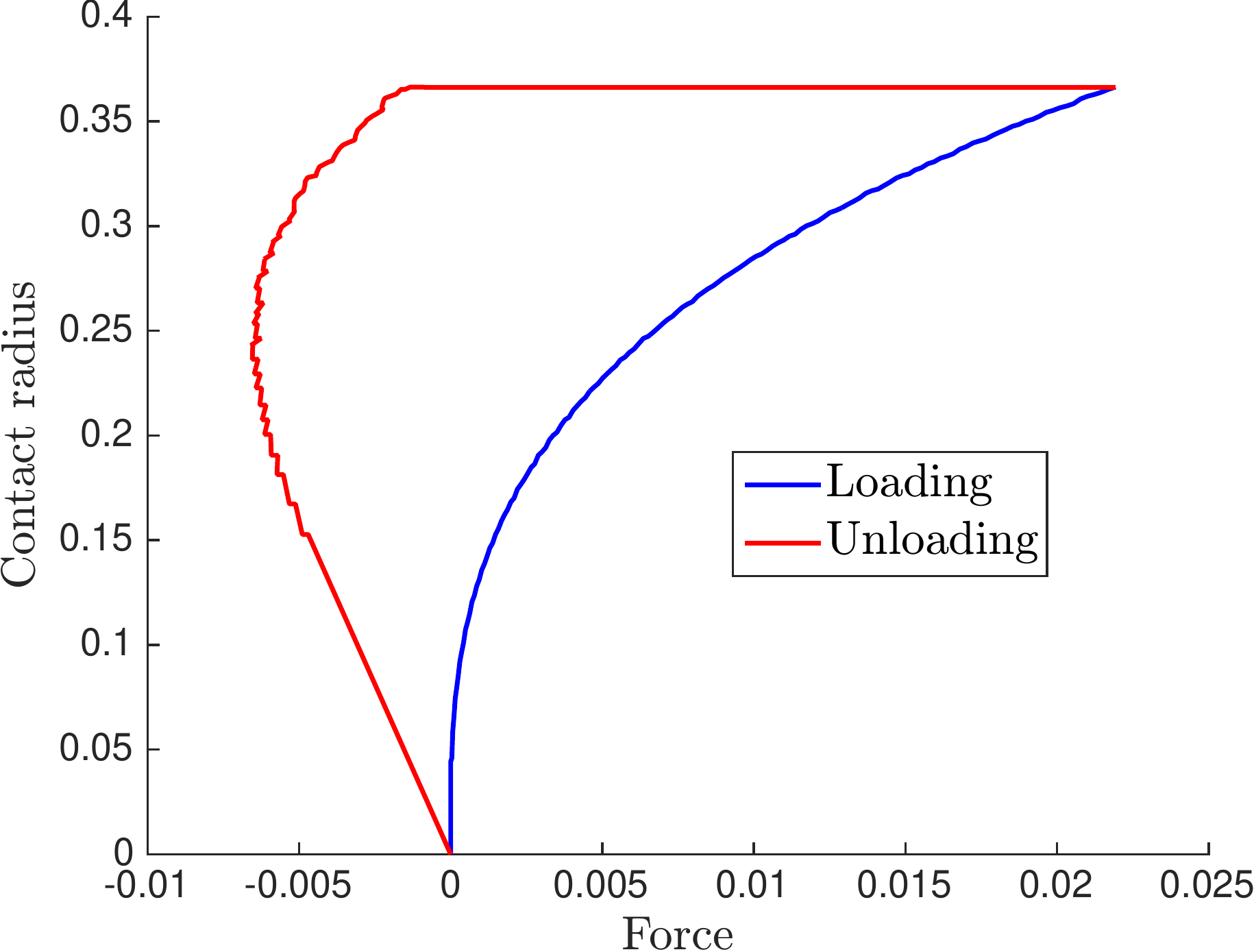}
\subcaption{}
\label{fig:forceContactRadiusLoadingUnloading128by128ScalingPower0}
\end{subfigure} 
\caption{Loading-unloading test of an adhesive sphere against a rigid flat
surface, variation of (a) force with indentation depth and (b) contact radius
with force. Since the springs can sustain tension up to a threshold force, the
force during unloading is different from that during loading. As we start
pulling the surfaces apart, the force becomes tensile, the springs start
breaking contact (wiggles in the curves), the force goes through a tensile
maximum and eventually drops to zero. During unloading, the area initially
remains constant and then starts decreasing as the springs snap off. The
force and area drop to zero from a finite nonzero value as observed in some
experiments \cite{johnsonKL:1}.}
\label{fig:sphereLoadingUnloading}
\end{figure}

The loading-unloading behavior is shown in Figure
\ref{fig:sphereLoadingUnloading}. During the loading phase (shown in blue), all
springs are in compression and the system is exactly the same as in Hertzian
contact. However, the unloading behavior (shown in red) is different and more
interesting.  Some of the springs are now in tension and thus at the same
indentation depth, the total force is different from that during loading. The
unloading curve is always lower (force more tensile) than the loading curve. As
the surfaces are pulled apart, the springs that reach their threshold force
break out of contact. This can be seen as wiggles in the unloading part of the
curve (and they are present only in the unloading phase). Upon continued
unloading, the total force reaches a tensile maximum.  Since we are performing
a displacement-controlled simulation, upon further unloading, the tensile force
decreases from the maximum and drops to $0$ as all springs snap off (Figure
\ref{fig:dilatationForceLoadingUnloading128by128ScalingPower0}). If we did a
force-controlled test, the entire contact would break and the force drop to
zero at the maximum tensile force.  During unloading, before the first spring
breaks contact, the contact area remains constant even as the force becomes
tensile (Figure
\ref{fig:forceContactRadiusLoadingUnloading128by128ScalingPower0}). As the
springs start snapping, the area gradually decreases and goes to zero.  Observe
that in the final step, both force and contact area drop to zero from a finite
nonzero value. This has been observed in experiments as well
\cite{johnsonKL:1}.

\subsection{Macroscopic behavior is discretization dependent}

\begin{figure}[H]
\centering
\begin{subfigure}[t]{0.49\textwidth}
\includegraphics[width=\textwidth]{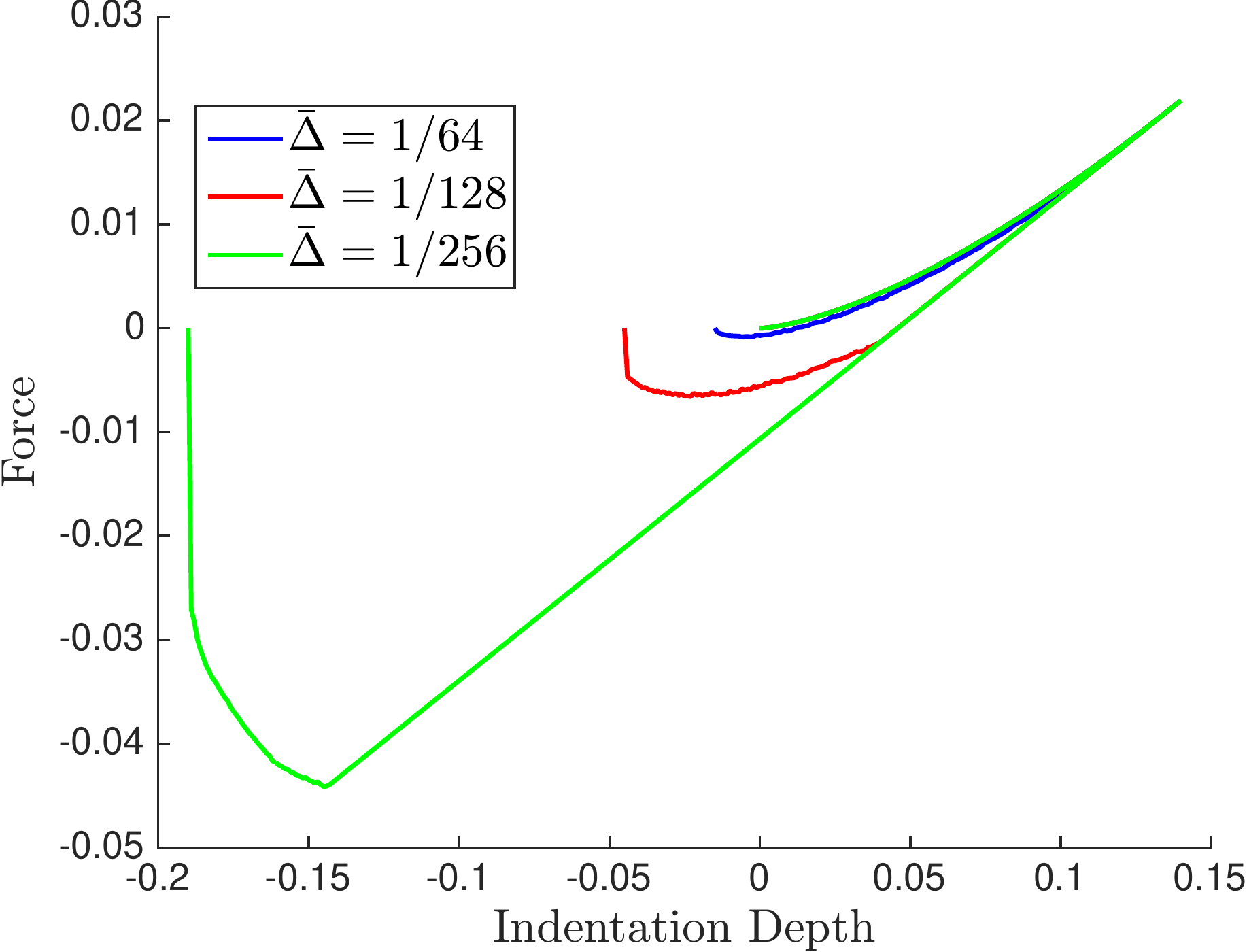}
\subcaption{}
\label{fig:dilatationForceDiscretizationDependenceScalingPower0}
\end{subfigure} 
\begin{subfigure}[t]{0.49\textwidth}
\includegraphics[width=\textwidth]{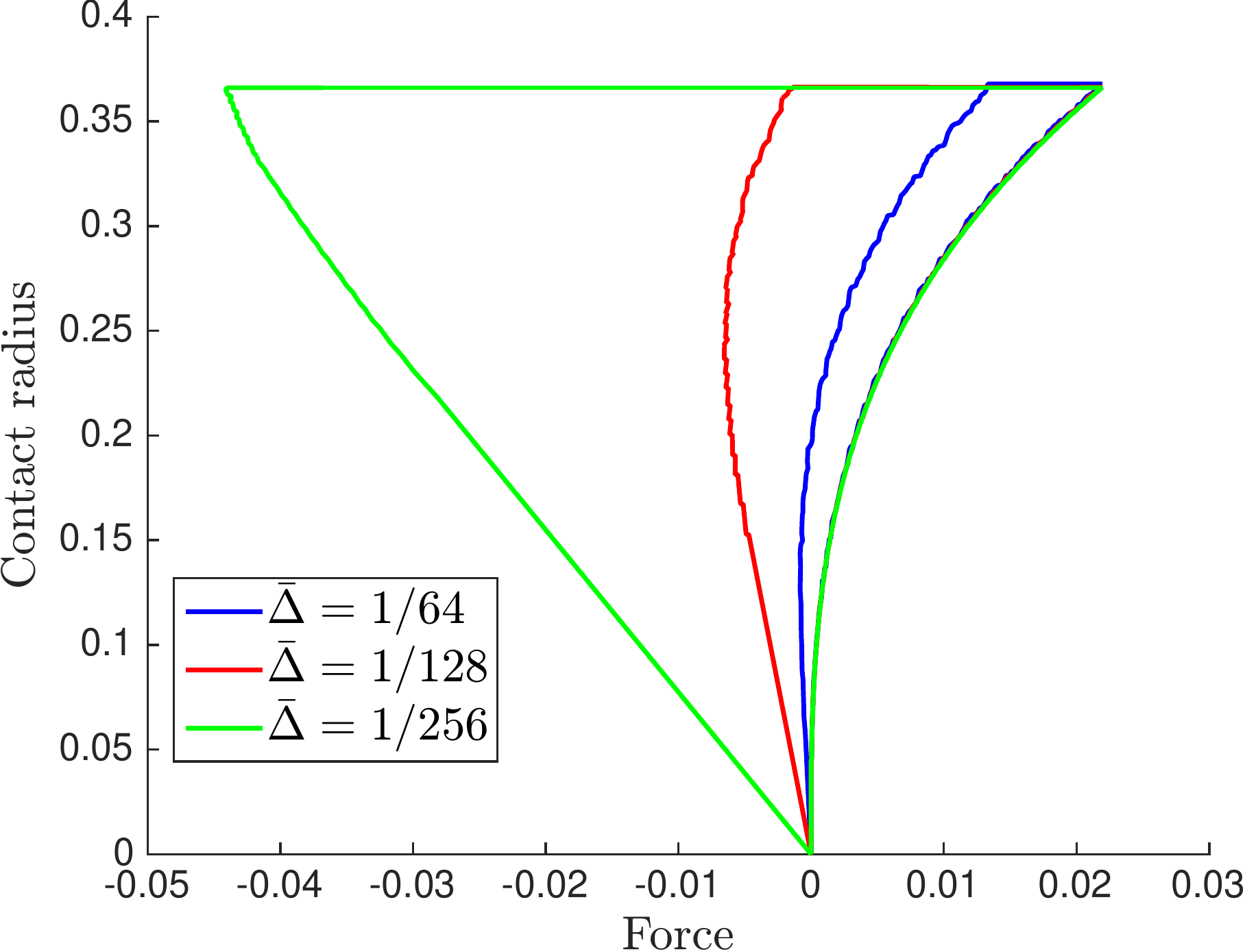}
\subcaption{}
\label{fig:forceContactRadiusDiscretizationDependenceScalingPower0.pdf}
\end{subfigure} 
\caption{For the same threshold force, the unloading response depends on the
discretization size. The results suggest that the threshold force must decrease
with decreasing $\bar{\Delta}$ but what is the right dependence for convergent
macroscopic behavior?}
\label{fig:discretizationDependenceScalingPower0}
\end{figure}

Let us study how the numerical solution behaves as we discretize the
sphere with more and more springs. We repeat the above simulations with
$\bar{\Delta} = 1/64, 1/128, 1/256$ (radius of sphere = 1, $\bar{F}_{th} =
2\times 10^{-5}$). The response in the loading phase is the same as that in
Hertzian contact (Section \ref{subsec:hertzianContactValidation}).  During
unloading, the behavior turns out to be discretization dependent (Figure
\ref{fig:discretizationDependenceScalingPower0}).  The maximum tensile force
increases with decreasing $\bar{\Delta}$ (Figure
\ref{fig:dilatationForceDiscretizationDependenceScalingPower0}).  The existence
of dependence on discretization only during unloading suggests that $\bar{F}_{th}$
cannot be independent of the discretization size; it must decrease with
decreasing $\bar{\Delta}$ for convergent macroscopic behavior. The question is,
how should $\bar{F}_{th}$ vary with $\bar{\Delta}$?

\subsection{A scaling suggestive of Linear Elastic Fracture Mechanics (LEFM)}
\label{subsec:scalingLEFM}

\begin{figure}
\centering
\begin{subfigure}[t]{0.49\textwidth}
\includegraphics[width=\textwidth]{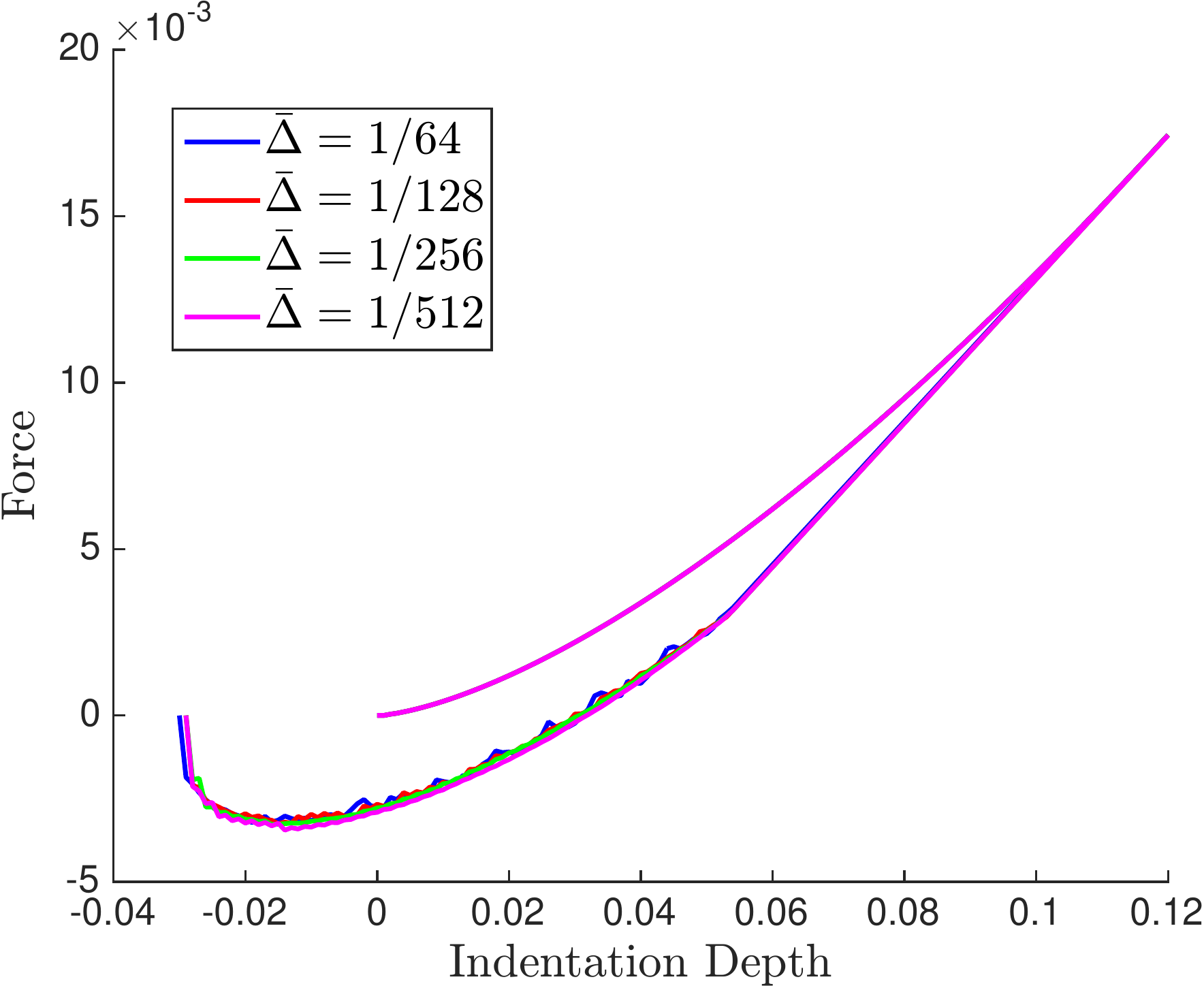}
\subcaption{}
\label{fig:dilatationForceDiscretizationDependenceScalingPower075}
\end{subfigure} 
\begin{subfigure}[t]{0.49\textwidth}
\includegraphics[width=\textwidth]{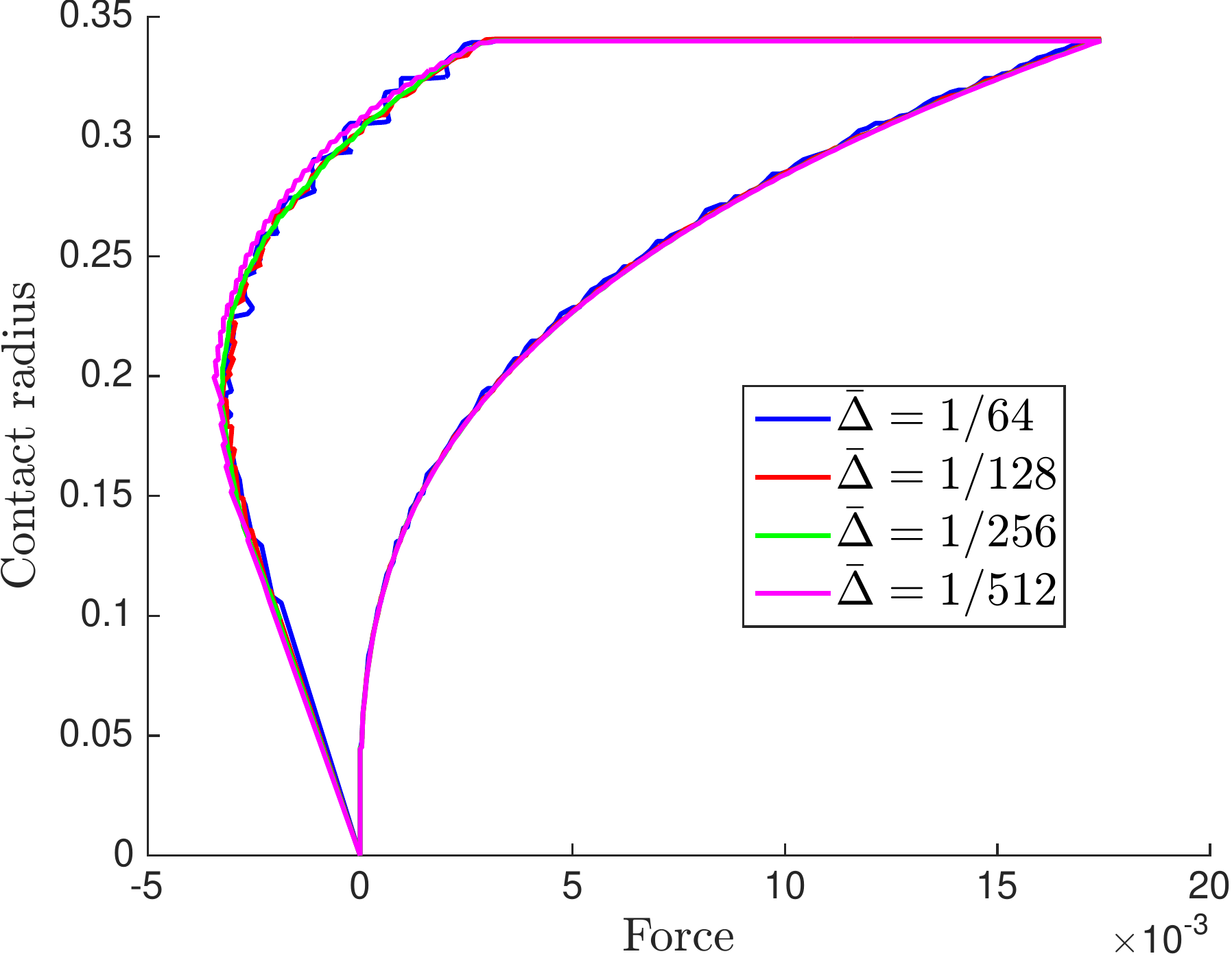}
\subcaption{}
\label{fig:forceContactRadiusDiscretizationDependenceScalingPower075.pdf}
\end{subfigure} 
\caption{If the threshold force is scaled as $\bar{F}_{th} = \bar{K}_{th}
A^{0.75}$ where $A = \bar{\Delta}^2$, the macroscopic behavior becomes
independent of $\bar{\Delta}$ (in the above simulations, radius of sphere = 1,
$\bar{K}_{th} = 2\times 10^{-2}$). This scaling turns out to be analogous to
the scaling of the critical stress near a crack tip in Linear Elastic Fracture
Mechanics!}
\label{fig:discretizationDependenceScalingPower075}
\end{figure}

For the dependence of threshold force on the discretization size, we consider
scalings of the form 

\begin{equation}\label{eq:thresholdForceScaling}
\bar{F}_{th} = \bar{K}_{th} A^\alpha,
\end{equation}
where $A = \bar{\Delta}^2$ is the area of the contact represented by one
spring, and $\bar{K}_{th}$ is a constant. Numerical experiments suggest that,
for $\alpha = 0.75$, the macroscopic behavior converges with decreasing
$\bar{\Delta}$ (Figure \ref{fig:discretizationDependenceScalingPower075}).

This appears puzzling at first, but a little calculation reveals an 
interesting connection with LEFM. The threshold force scales as 
\begin{equation}
\bar{F}_{th} = \bar{K}_{th} A^{0.75} \implies \sigma_{th} = \bar{F}_{th}/A
= \bar{K}_{th} A^{-0.25}
\end{equation}
where $\sigma_{th}$ is the threshold stress. Using
$A = \bar{\Delta}^2$, 
\begin{equation}\label{eq:scalingLEFM}
\sigma_{th} = \frac{K_{th}}{\sqrt{\bar{\Delta}}}.
\end{equation}
This scaling is suggestive of LEFM.  There, the stress at the crack tip is
given by (for mode I):
\begin{equation}\label{eq:stressLEFM}
\sigma \propto \frac{K_I}{\sqrt{r}},
\end{equation}
where $K_I$ is the mode I stress intensity factor and $r$ is the distance from
the crack tip.  The criterion for crack propagation is that the stress
intensity factor at the crack tip should equal the critical stress intensity
factor ($K_{IC}$),
\begin{equation}\label{eq:stressLEFM2}
K_I = K_{IC} \implies \sigma_C \propto \frac{K_{IC}}{\sqrt{r}},
\end{equation}
where $\sigma_C$ is the critical stress (singular at the crack tip). Comparing
equations (\ref{eq:scalingLEFM}) and (\ref{eq:stressLEFM2}), it is evident
that, with $\alpha = 0.75$, we have a scaling similar to LEFM with
$\bar{K}_{th}$ serving the role of $K_{IC}$. Adhesive contact can be seen from
the perspective of fracture mechanics with the area not in contact interpreted
as an external crack \cite{maugis1992adhesion}. This also explains the
discretization-size dependence we saw in Figure
\ref{fig:discretizationDependenceScalingPower0}. Without the $\alpha = 0.75$
scaling (there $\alpha = 0$), by changing $\bar{\Delta}$, we effectively
changed $\bar{K}_{th}$ (or $K_{IC}$). 

To further bolster this connection between the scaling (Equation
(\ref{eq:scalingLEFM})) and LEFM, we performed two kinds of tests. First, we
did simulations with various other geometries and in each case found that the
macroscopic response is discretization-size independent for $\alpha = 0.75$ and
dependent for $\alpha \neq 0.75$. Further, in our model, the elasticity is
captured in the kernel $1/\bar{r}_{ij}$. If the scaling is related to LEFM, it
must depend on this kernel. Simulations with other kernels, for example with
only local interaction ($\bar{C}_{ij} = C_0 \delta_{ij}$) or with only a few
nearest-neighbor interactions, show that the scaling indeed depends on the
kernel. When $\bar{C}_{ij} = C_0 \delta_{ij}$, i.e., when there is no elastic
interaction, the macroscopic response is $\bar{\Delta}$-dependent for $\alpha =
0.75$ and independent for $\alpha = 1$. 

\section{Can the model reproduce the JKR results?} \label{sec:JKR}

Our model is JKR-like since the tensile forces act only within the contact area
(as opposed to the DMT model where tensile forces act even outside the
contact). Further, the JKR model can also be seen from the perspective of LEFM
\cite{maugis1992adhesion}. Let us further compare the two models. In the JKR
model, the maximum tensile force $F_{\text{max}}$ is given by
\begin{equation}\label{eq:adhesionStrengthJKR} F_{\text{max}} = \frac{3}{2} \pi
\gamma R \end{equation} where $\gamma$ is the fracture energy and $R$ is the
radius of the sphere. The maximum force is linear in both the fracture energy
and the radius. Let us check these scalings in our model.

\begin{figure}[H]
\centering
\begin{subfigure}[t]{0.49\textwidth}
\includegraphics[width=\textwidth]{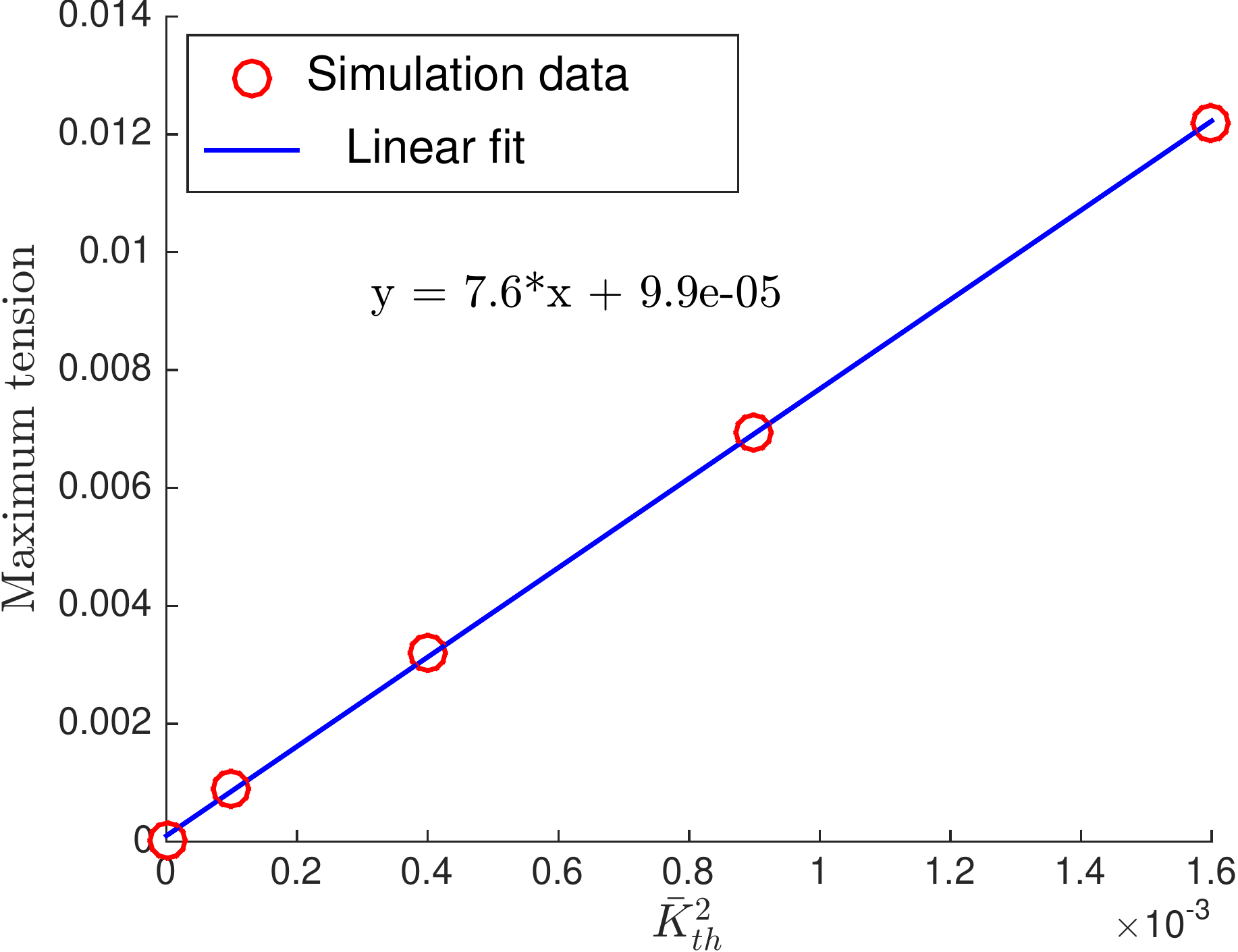}
\subcaption{}
\label{fig:thresholdForceMaxTensionScaling}
\end{subfigure} 
\begin{subfigure}[t]{0.49\textwidth}
\includegraphics[width=\textwidth]{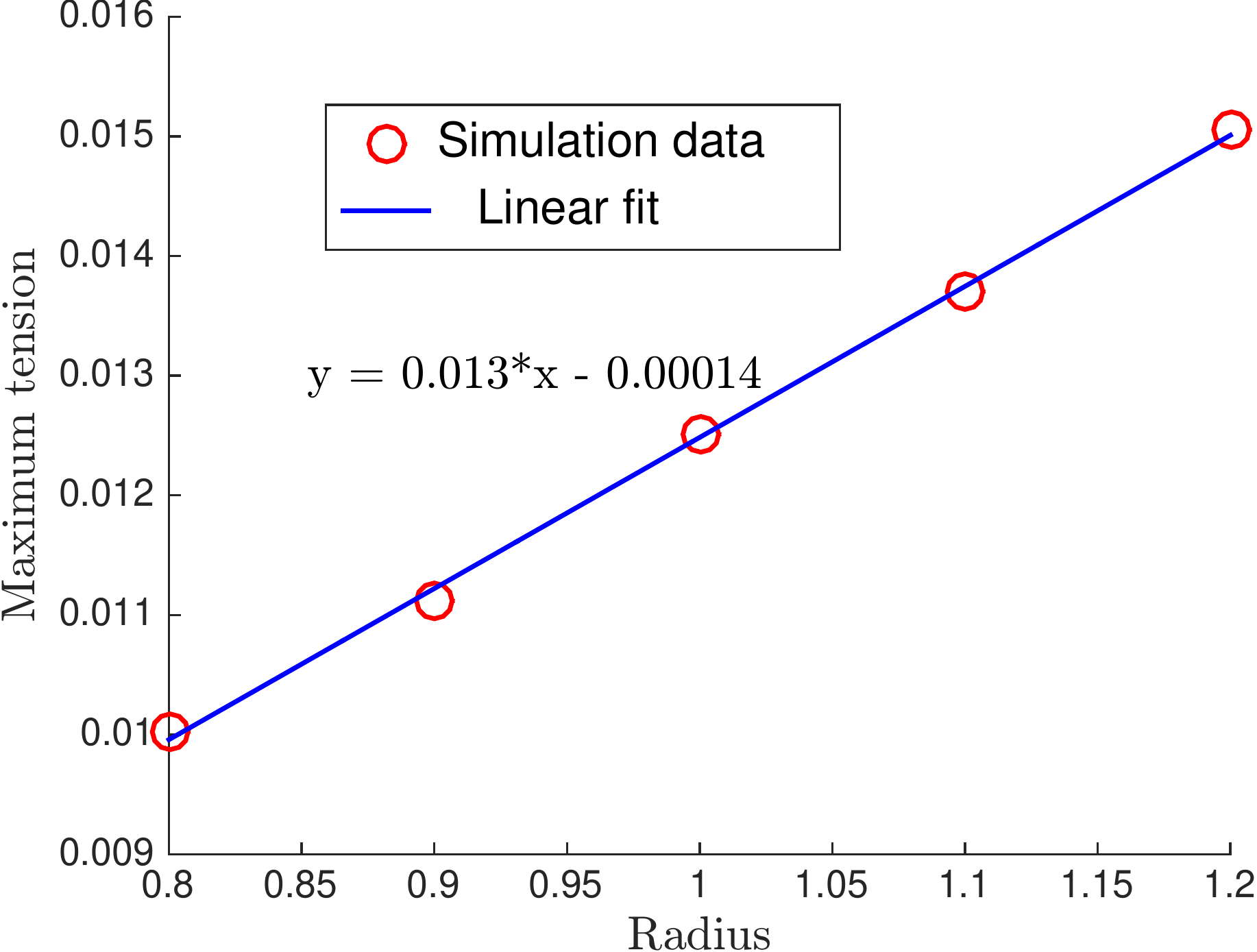}
\subcaption{}
\label{fig:radiusMaxTensionScaling}
\end{subfigure} 
\caption{Scaling of the maximum tensile force with the (a) square of the
parameter $\bar{K}_{th}$ and (b) radius. These scalings are the same as in
the JKR model.}
\label{fig:thresholdForceRadiusMaxTensionScaling}
\end{figure}

We saw that in our model, $\bar{K}_{th}$ is like the critical stress
intensity factor $K_{IC}$ in fracture mechanics (Equations
(\ref{eq:scalingLEFM}) and (\ref{eq:stressLEFM2})). In LEFM, the critical
energy release rate for mode I is $G_{IC} \propto K_{IC}^2$.  If this is the
fracture energy $\gamma$, from Equation (\ref{eq:adhesionStrengthJKR}), and if the
same scaling is to hold in our model, the maximum tensile force
$\bar{F}_{\text{max}}$ must scale as $\bar{K}_{th}^2$. To check this, we
repeated the simulations described earlier, increasing $\bar{K}_{th}$ from 0
to 0.04 in steps of 0.01 (radius of sphere = 1, $\bar{\Delta} = 1/256$).  We
find that $\bar{F}_{\text{max}}$ does scale as $(\bar{K}_{th})^2$ (Figure
\ref{fig:thresholdForceMaxTensionScaling}).

Further, $\bar{F}_{\text{max}}$ must increase linearly with the radius
(Equation (\ref{eq:adhesionStrengthJKR})). This is verified in Figure
\ref{fig:radiusMaxTensionScaling} where we show how the maximum tension varies
with radius $(\bar{K}_{th} = 0.04, \bar{\Delta} = 1/256)$.

Our numerical experiments indicate that (Figure
\ref{fig:thresholdForceRadiusMaxTensionScaling}):  
\begin{equation}\label{eq:adhesionStrengthModel}
\bar{F}_{\text{max}} = 7.6 \bar{K}_{th}^2 \bar{R}
\implies F_{\text{max}} = \frac{7.6(1-\nu)}{2\pi G}K_{th}^2 R
\end{equation}
where $K_{th}$ has units of  MPa $\sqrt{m}$, the same as that of $K_{IC}$
(see the nondimensionalization discussion). From equations 
\ref{eq:adhesionStrengthJKR} and \ref{eq:adhesionStrengthModel},
\begin{equation}\label{eq:thresholdForceParameterDependence}
\frac{7.6(1-\nu)}{2\pi G}K_{th}^2 = \frac{3\pi\gamma}{2}
\implies K_{th} = \pi \sqrt{\frac{3G\gamma}{7.6(1-\nu)}}.
\end{equation}
This gives the threshold-force parameter $K_{th}$ in our model in
terms of the fracture energy $\gamma$ and elastic properties $G,\nu$.  Note that
$K_{th}$ is a material property independent of the geometry. If $\gamma, G,
\nu$ of a material are known, $K_{th}$ can be determined using Equation
(\ref{eq:thresholdForceParameterDependence}) and used in simulations for any
surface geometry.

\section{Conclusion}\label{sec:conclusion}

In a system of discrete springs, we study a threshold-force based model for
adhesive contact and mode I fracture. Using a scaling of the threshold force
with discretization size, we demonstrate the analogy between our model,
fracture mechanics, and the JKR adhesion model.  The advantage of the model
presented here is that it is conceptually simple and easy to implement
numerically. Possible applications of the model include studying rough-surface
adhesion with long-range elastic interactions and calculation of stress
intensity factors for various geometries for mode I fracture. Extension to
modes II and III might be possible by adding horizontal degrees of freedom to
the springs and using an appropriate kernel based on elasticity solutions.
Time-dependent behavior such as viscoelasticity can be incorporated by making
the interaction kernel time-dependent ($C_{ij}(t)$) based on solutions of a
viscoelastic half-space boundary value problem. In this case, some of the
elastic energy stored will be lost by viscoelastic dissipation and interesting
rate effects might emerge.

\section*{Acknowledgment}
We gratefully acknowledge the support for this study from the National Science
Foundation (grant EAR 1142183) and the Terrestrial Hazards Observations and
Reporting center (THOR) at Caltech. 

\bibliographystyle{unsrt}
\bibliography{./criticalTensileForceModel}

\end{document}